\documentclass[letterpaper, 10 pt, conference]{ieeeconf}  %
\IEEEoverridecommandlockouts                              %
\usepackage{amsmath} %
\usepackage{amssymb}  %
\usepackage{mathtools}
\usepackage{graphicx}
\usepackage{color}
\usepackage{xcolor}
\usepackage{cite}

\usepackage[hyphens]{url}	%

\usepackage{siunitx}
\usepackage{tikz}
\usetikzlibrary{
    angles,
    arrows,
    arrows.meta,
    automata,
    calc,
    fit,
    patterns,
    positioning,
    quotes,
    shapes,
}
\tikzset{pill/.style={minimum width=1.2cm,minimum height=6mm,rounded
corners=3mm,draw}}
\tikzstyle{block} = [draw, thick, node distance=1cm, inner sep=6pt, minimum height=1cm, minimum width=1.7cm, align=center, fill=white]
\usepackage{pgfplots}
\pgfplotsset{compat=newest}
\usepgfplotslibrary{external}

\makeatletter
\let\NAT@parse\undefined
\makeatother
\usepackage[hidelinks]{hyperref} %

\newcommand\copyrighttext{%
    \footnotesize
    \textcopyright 2025 IEEE. Personal use of this material is permitted.
    Permission from IEEE must be obtained for all other uses, in any current or future media, including reprinting/republishing this material for advertising or promotional purposes, creating new collective works, for resale or redistribution to servers or lists, or reuse of any copyrighted component of this work in other works.
}
\newcommand\copyrightnotice{%
    \begin{tikzpicture}[remember picture,overlay]
        \node[anchor=south,yshift=10pt] at (current page.south) {\fbox{\parbox{\dimexpr\textwidth-\fboxsep-\fboxrule\relax}{\copyrighttext}}};
    \end{tikzpicture}%
}

\title{\LARGE \bf
Path-following model predictive control for autonomous e-scooters
}

\author{David Meister$^1$, Robin Strässer$^1$, Felix Brändle$^1$, Marc Seidel$^1$, Benno Bassler$^2$, Nathan Gerber$^2$,\\Jan Kautz$^2$, Elena Rommel$^2$, Frank Allgöwer$^1$%
\thanks{F.\ Allgöwer is thankful that this work was funded by the Ministry of Science, Research and the Arts of the State of Baden-Württemberg (MWK) in the context of the ``MobiLab'' Project.
D.\ Meister, R.\ Strässer, M.\ Seidel thank the Graduate Academy of the SC SimTech for its support.
F.\ Brändle thanks the International Max Planck Research School for Intelligent Systems (IMPRS-IS) for its support.}%
\thanks{$^{1}$D.\ Meister, R.\ Strässer, F.\ Brändle, M.\ Seidel, F.\ Allgöwer are with the University of Stuttgart, Institute for Systems Theory and Automatic Control, 70550 Stuttgart, Germany
(e-mail: e-scooter@ist.uni-stuttgart.de).}%
\thanks{$^{2}$B.\ Bassler, N.\ Gerber, J.\ Kautz, E.\ Rommel are students at the University of Stuttgart, 70550 Stuttgart, Germany.}%
}

\begin{document}
\maketitle
\thispagestyle{empty}
\pagestyle{empty}

\begin{abstract}
    In order to mitigate economical, ecological, and societal challenges in electric scooter (e-scooter) sharing systems, we develop an autonomous e-scooter prototype.
    Our vision is to design a fully autonomous prototype that can find its way to the next parking spot, high-demand area, or charging station.
    In this work, we propose a path-following model predictive control solution to enable localization and navigation in an urban environment with a provided path to follow.
    We design a closed-loop architecture that solves the localization and path following problem while allowing the e-scooter to maintain its balance with a previously developed reaction wheel mechanism.
    Our model predictive control approach facilitates state and input constraints, e.g., adhering to the path width, while remaining executable on a Raspberry Pi 5.
    We demonstrate the efficacy of our approach in a real-world experiment on our prototype.
\end{abstract}


\section{Introduction}

Electric scooter (e-scooter) sharing systems are deployed in many cities worldwide~\cite{Goessling2020}.
They are often used in urban mobility, in particular for short-distance commutes~\cite{Degele2018}.
In this context, free-floating sharing systems allow for greater operational flexibility compared to station-based systems.
While this has many advantages for the user, it also causes practical challenges such as obstructed public spaces due to improperly parked or dropped e-scooters, or increased relocation efforts.
To address these issues, suppliers often rely on CO$_2$- and labor-intensive methods to relocate e-scooters.
Moreover, high availability of e-scooters typically involves many e-scooters to be deployed in operation areas.
These factors result in economical, ecological, and social costs of current e-scooter sharing systems~\cite{Heineke2019,Hollingsworth2019,Cazzola2020,Gioldasis2021}.
\copyrightnotice%

To mitigate many of these concerns, our research group has proposed the development of autonomous e-scooters that can self-stabilize, navigate, and avoid collisions~\cite{Wenzelburger2020,Soloperto2021,strasser:seidel:braendle:meister:soloperto:hambach-ferrer:allgoewer:2024,Straesser2025,brandle:meister:seidel:strasser:allgower:2025}.
An autonomous e-scooter is a standard two-wheeled e-scooter equipped with a reaction wheel to balance and additional hardware to enable autonomous operation in pedestrian environments while not being used by a human rider.
This eliminates the need for manual redistribution or collection for charging, which can be costly and inefficient.
This autonomous micromobility solution allows us to combine the advantages of station-based and free-floating sharing systems:
With an autonomous e-scooter, riders can end their journey anywhere within a designated area.
In contrast to conventional e-scooters in free-floating sharing systems, the autonomous e-scooter can autonomously park itself after use, avoiding cluttered public spaces.
Moreover, autonomous e-scooters can optimize their distribution over an operation area according to current or forecasted demand.
This improves availability and reduces the number of e-scooters required to facilitate an operation area.
This approach has the potential to make e-scooter sharing systems more sustainable and efficient.

\emph{Related work:}
Related to our work on balancing autonomous e-scooters~\cite{Wenzelburger2020,Soloperto2021}, the authors in~\cite{lin:jafari:liu:2024} describe how riderless e-scooters can maintain their balance.
The authors propose a Proportional-Derivative (PD) controller and a feedback-linearized PD controller and evaluate their functionality in simulation.
In \cite{Poojari2024}, another autonomous e-scooter prototype is presented which uses support wheels instead of a balancing mechanism.
The authors describe a path following algorithm based on a Timed-Elastic-Band Optimization Problem.
Since our e-scooter does not rely on support wheels but uses a reaction wheel to stabilize the roll dynamics equilibrium, we propose a model predictive control (MPC) approach for path following that also accounts for the balancing dynamics.
In particular, autonomous driving while sliding along the roll dynamics equilibria further constrains the online motion planner.
Hence, we not only account for the non-holonomic dynamics of the e-scooter, as \cite{Poojari2024}, but also design constraints that only allow for motion plans permitting stabilization in the roll dynamics as well.

As a prerequisite for path following, we solve the localization problem, identifying the e-scooter position and orientation.
The Global Navigation Satellite System (GNSS) is a crucial technology for this task.
An overview on state estimation approaches for navigation of automated vehicles can be found in \cite{Konrad2018}.
The authors differentiate between loosely- and tightly-coupled GNSS-based localization.
The former determines the global position purely based on GNSS measurements and combines them a posteriori with other onboard sensors such as encoders or an Inertial Measurement Unit (IMU), e.g., \cite{Wendel2006}.
A tightly-coupled integration avoids computing a purely GNSS-based position estimate and directly combines raw GNSS measurements with onboard sensors, e.g., \cite{Godha2007}.
We rely on a loosely-coupled localization approach in this work which is based on an Extended Kalman Filter (EKF) that combines processed GNSS measurements with encoder measurements and motion dynamics.

In order to enable the autonomous e-scooter to move from one location to another without a rider, we solve a path following problem.
Balancing and local path planning require taking the system dynamics and constraints into account.
The research community has proposed various MPC approaches addressing these challenges.
For example,~\cite{maedeh:cobbenhagen:sommer:andrien:lefeber:heemels:2024} propose a high-performance MPC for quadcopters with stability guarantees.
It is also important to differentiate between trajectory tracking and path following schemes.
While the former, e.g.,~\cite{Limon2008}, specify reference points over time, the latter, e.g.,~\cite{Faulwasser2009}, relax the problem and follow a parametrized path along which the controller finds a time discretization online.
We refer to~\cite{Stano2023} for a survey on path following approaches for automated road vehicles.
However, these vehicles usually come with four wheels, not requiring a balancing solution.
In this work, we assume to be provided with a path to follow which may violate the dynamics of the e-scooter.
Hence, we do not prescribe a reference trajectory and the e-scooter needs to find a discretization of the path online. %

\emph{Contribution:}
In this paper, we propose a path following approach for the autonomous e-scooter shown in Fig.~\ref{fig:scooter_labeled} using an MPC scheme and leveraging a localization solution based on an EKF.
A particular challenge in contrast to the literature is the combination of a path following problem with a balancing task for our autonomous e-scooter prototype.
We propose a closed-loop architecture that allows us to address both challenges simultaneously and can still be run on a Raspberry Pi 5 and the motor controllers required for actuation.
Our proposed solution addresses the path following problem by locally identifying a time discretization for the path, thereby turning it locally into a reference trajectory, and then applying a reference tracking MPC scheme.
We demonstrate the efficacy of our proposed control architecture in a real-world experiment on our prototype.\footnote{\label{foot:video}A video of our experiment can be found at: \href{https://www.ist.uni-stuttgart.de/research/Files/MPC-Path-Following.mp4}{https://www.ist.uni-stuttgart.de/research/Files/MPC-Path-Following.mp4}. More video material on our project including more test runs can be found at: \href{https://www.estarling.io}{estarling.io}.}

\emph{Outline:}
In Section~\ref{sec:setup}, we describe our prototype and problem setup.
The proposed closed-loop architecture for solving the path following problem for our prototype is presented in Section~\ref{sec:cl_arch}.
Building on this architectural design, we present our localization and path following scheme, based on an EKF and MPC approach, in Sections~\ref{sec:ekf} and~\ref{sec:mpc}.
We demonstrate the effectiveness of our proposed solution in a real-world experiment on our prototype in Section~\ref{sec:exp}, before concluding our paper in Section~\ref{sec:concl}.

\section{Setup}
\label{sec:setup}

In this section, we describe our autonomous e-scooter prototype and the problem setup.
As explained in the introduction, our goal is to enable an e-scooter to follow a pre-defined path along GNSS coordinates and the connecting line segments.
In the final application, this feature will be needed in order to relocate the autonomous e-scooters from one location to another along a grid of accessible paths.
In this paper, we assume that such a path defined by GNSS coordinates and the connecting line segments is provided by an external path generator, including path width information.
Our goal is to follow this path given the dynamics of our hardware and the path width constraints, requiring us to locate the e-scooter in the world frame.

\begin{figure}
    \centering
    \input{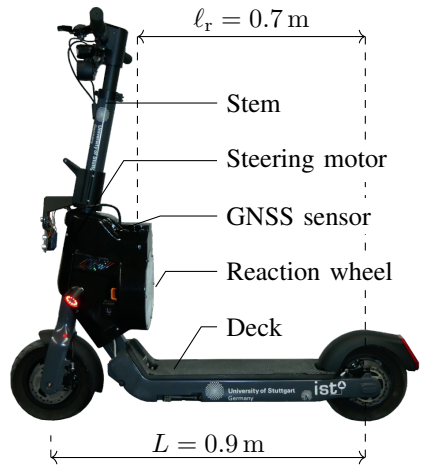}
    \caption{Our prototype with reaction wheel and GNSS sensor.}
    \label{fig:scooter_labeled}
    \vspace*{-\baselineskip}
\end{figure}

Our prototype consists of a commercially available e-scooter that has been equipped with an actuated reaction wheel and two IMUs in order to balance, i.e., keeping the e-scooter upright without a rider.
More details on the balancing mechanism can be found in~\cite{Wenzelburger2020} and an explanation on why two IMUs are needed for this balancing mechanism is provided in~\cite{brandle:meister:seidel:strasser:allgower:2025}.
The e-scooter has a wheel axle distance of $L=\SI{0.9}{\meter}$ and is furthermore equipped with a GNSS sensor consisting of an antenna and a ublox ZED-F9P module.
Moreover, we utilize Real-Time Kinematic (RTK) positioning via the SAPOS-HEPS service, incorporating correction data from approximately 270 GNSS reference stations in Germany, for improved accuracy.
We obtain GNSS measurements at approximately $\SI{10}{\hertz}$.
Our high-level planning algorithms proposed in this work will be executed on a Raspberry Pi 5 while low-level control of the motors is done on motor controllers, e.g., VESCs.
Some important components of our prototype are highlighted in Fig.~\ref{fig:scooter_labeled}.

The nonholonomic dynamics of the e-scooter, due to the limited steering angle and no immediate lateral motion, require preplanning of future system trajectories to solve the desired path following task.
Moreover, this requires soft violation of the prescribed path which consists of line segments and cannot be followed while adhering to the system dynamics, including also the balancing dynamics.
With these challenges in mind, we propose a closed-loop architecture consisting mainly of an EKF state estimator for localization of the e-scooter and an MPC scheme for path following.
We elaborate the details of the closed-loop architecture in the next section.

\section{Proposed closed-loop architecture}
\label{sec:cl_arch}

In order to achieve the goal of path following, we propose the following architecture of the closed loop, also depicted in Fig.~\ref{fig:cl_arch}.
\begin{figure}
    \centering
    \input{img/controller_structure.tex}
    \caption{Proposed closed-loop architecture. The main focus of this work are the components in the high-level control box.}
    \label{fig:cl_arch}
    \vspace*{-\baselineskip}
\end{figure}
The e-scooter localization is enabled via an EKF that combines GNSS measurements $z$ and the e-scooter dynamics, i.e., a kinematic single track model with encoder measurements, yielding state estimate $\hat{x}_\mathrm{EKF}$.
As a path following controller, we employ an MPC algorithm that predicts the e-scooter dynamics and determines the control inputs, i.e., the commanded velocity $v_\mathrm{cmd}$ and the commanded steering angle $\delta_\mathrm{cmd}$, for the low-level controllers.
As an additional input, the path following controller receives a path to follow from an external path generator.
Together, localization and path following controller are referred to as the high-level control layer in this work.
The low-level controllers comprise a steering stepper motor controller (Proportional-Integral (PI)-controller), a driving velocity controller (PI-controller), and the balancing controller actuating the reaction wheel as described in~\cite{Wenzelburger2020,Soloperto2021}.
These low-level controllers apply torques $\tau_\mathrm{str}, \tau_\mathrm{vel}, \tau_\mathrm{bal}$ to the e-scooter through the installed motors.
For curve driving, the balancing controller also filters the commanded velocity and steering angle such that balancing while turning can be maintained, yielding $\tilde{v}_\mathrm{cmd}, \tilde{\delta}_\mathrm{cmd}$.
We do not provide details on the deployed filtering method in this paper, but conservatively abstract this low-level filtering in the constraints of the MPC scheme used as path following controller.
Apart from this required dependency between high- and low-level control, our approach relies on fast low-level controllers in order to neglect the low-level tracking dynamics in the high-level control layer.
This also includes a feed-forward component in the driving controller which we are developing and can account for sloped terrain.
Moreover, the low-level controllers compensate disturbances from sources like wind and uneven terrain.

Note that the proposed architecture is fully compatible with the safety filter from our previous work on collision avoidance with ultrasonic sensors~\cite{strasser:seidel:braendle:meister:soloperto:hambach-ferrer:allgoewer:2024}.
This allows us to complement the presented path following scheme with an additional safety feature and demonstrates the power of the proposed safety filter approach in~\cite{strasser:seidel:braendle:meister:soloperto:hambach-ferrer:allgoewer:2024}.

\section{Localization as state estimation problem}
\label{sec:ekf}

\begin{figure}
    \centering
    \resizebox{0.9\columnwidth}{!}{%
        \input{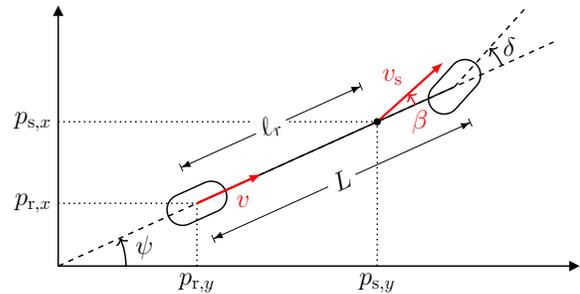}
    }
    \caption{Single track model variables.}
    \label{fig:variables_model}
    \vspace*{-\baselineskip}
\end{figure}

For high-level planning, we are required to localize the e-scooter, i.e., identify its position and orientation in the world frame.
To this end, we employ a GNSS sensor for global positioning, but reduce measurement noise and estimate the e-scooter orientation by combining the measurements via the kinematic single track model
\begin{equation}\label{eq:single_track_model}
    x_\mathrm{EKF}
    = \begin{bmatrix} p_{\mathrm{s},x} \\ p_{\mathrm{s},y} \\ \psi \end{bmatrix},
    \quad
    \dot{x}_\mathrm{EKF}
    = \begin{bmatrix} v_\mathrm{s} \cos(\psi+\beta) \\ v_\mathrm{s} \sin(\psi+\beta) \\ v \tan(\delta) / L \end{bmatrix},
\end{equation}
where $p_{\mathrm{s},x}, p_{\mathrm{s},y}$ specify the position of the e-scooter (more precisely, of the GNSS sensor), $\psi$ denotes the e-scooter's orientation, $v$ is its velocity at the rear axle, $\delta$ is the steering angle, and $L$ refers to the wheel axle distance.
We abbreviated the side-slip angle at the GNSS sensor mounting point $\beta = \arctan(\ell_\mathrm{r} \tan(\delta)/L)$ and the absolute velocity of the e-scooter at the GNSS sensor mounting point $v_\mathrm{s} = v \sqrt{1 + \tan(\beta)^2}$.
The introduced variables are depicted in Fig.~\ref{fig:variables_model}.
We utilize Euler discretization in order to transform the continuous-time model into a discrete-time one for implementation.
The current steering angle $\delta$ and velocity $v$ are obtained from encoder measurements.

The chosen position coordinates $p_{\mathrm{s},x}, p_{\mathrm{s},y}$ at the GNSS sensor mounting point allow for a linear measurement model
\begin{equation*}
    z = \begin{bmatrix} p_{\mathrm{s},x} \\ p_{\mathrm{s},y} \end{bmatrix}
    = \begin{bmatrix} 1 & 0 & 0 \\ 0 & 1 & 0 \end{bmatrix} x_\mathrm{EKF}.
\end{equation*}
We assume additive process and measurement noise, where the covariance of the former is tuned and the one of the latter is provided online based on the quality of the received satellite signals and the RTK status by the employed GNSS processing board.

The prediction and correction step of the EKF are well-known, and we refer the reader to~\cite{Simon2006} for the mathematical details.
We run the EKF at $\SI{10}{\hertz}$ on our prototype.
Note that we can transform the state estimate $\hat{x}_\mathrm{EKF}$ to any point along the e-scooter via a static transformation, in particular, the rear and front wheel axle.
We can transform to the front axle position $p_{\mathrm{f},x}, p_{\mathrm{f},y}$ via
\begin{equation}\label{eq:pos_f_trafo}
    \begin{bmatrix} p_{\mathrm{f},x} \\ p_{\mathrm{f},y} \\ \psi \end{bmatrix}
    = \begin{bmatrix} p_{\mathrm{s},x} + (L-\ell_\mathrm{r}) \cos(\psi) \\ p_{\mathrm{s},y} + (L-\ell_\mathrm{r}) \sin(\psi)\\ \psi \end{bmatrix},
\end{equation}
and to the rear axle position $p_{\mathrm{r},x}, p_{\mathrm{r},y}$ via
\begin{equation}\label{eq:pos_r_trafo}
    \begin{bmatrix} p_{\mathrm{r},x} \\ p_{\mathrm{r},y} \\ \psi \end{bmatrix}
    = \begin{bmatrix} p_{\mathrm{s},x} - \ell_\mathrm{r} \cos(\psi) \\ p_{\mathrm{s},y} - \ell_\mathrm{r} \sin(\psi)\\ \psi \end{bmatrix}.
\end{equation}
Importantly, these transformations influence the state estimate covariance obtained in the EKF.
However, we do not process this covariance information in our MPC scheme for path following.

\section{Path following with an MPC scheme}
\label{sec:mpc}

In this section, we present an MPC scheme in order to solve the path following problem.
Throughout this section, we consider the e-scooter state
\begin{equation}
    x = \begin{bmatrix}
        p_{\mathrm{f},x} & p_{\mathrm{f},y} & v & \cos(\psi) & \sin(\psi) & \delta
    \end{bmatrix}^\top.
\end{equation}
We utilize a 2-step process to achieve our goal:
Firstly, we generate a local reference trajectory from the path to follow online.
The reference trajectory is called local because it only covers the portion of the path that is relevant for the next iteration of the path following MPC scheme.
Secondly, we solve an MPC reference tracking problem.
The following subsections contain detailed explanations on these steps.

\subsection{Discretization of the path into a trajectory}

We utilize discretization in order to turn our path following problem into a local trajectory tracking problem.
Note that there exist also direct path following approaches such as~\cite{Faulwasser2009} %
which we plan to explore in future research.
However, they are typically computationally more demanding in our setup and therefore pose additional challenges for application on our prototype.
Hence, in this work, we present a discretization approach to locally transform the path following problem into a trajectory tracking problem.

We start with a path defined by GNSS coordinates, called waypoints, and the line segments between consecutive waypoints.
Our approach discretizes a local portion of the path in time, yielding a reference trajectory for the MPC problem.
We consider an MPC prediction horizon of $T = \SI{6}{\meter} / v_\mathrm{max}$ with $v_\mathrm{max}=\SI{0.7}{\meter\per\second}$ in our application. %
Our MPC scheme runs with $f_\mathrm{MPC}=\SI{8}{\hertz}$.
Hence, we can discretize the prediction horizon into $N = T f_\mathrm{MPC}$ steps.
We define a look-ahead distance $d = 0.9 v_\mathrm{max} T$ of the MPC scheme along the remaining path to be followed.
Thus, we compute the maximal achievable distance along the path when driving with 90\% of full speed over the prediction horizon.
Considering the upcoming portion of the path of length $d$, we discretize into a reference trajectory by splitting this portion of the path into $N$ pieces of equal length in space and time, yielding $(p_{\mathrm{ref},x}(k),p_{\mathrm{ref},y}(k))_{k\in[0,N]}$.
The reference for the orientation $(\psi_\mathrm{ref}(k))_{k\in[0,N]}$ of the e-scooter is obtained for each of these points by pointing along the path in driving direction.
For the reference velocity, we use $v_\mathrm{ref}(k) = 0.9 v_\mathrm{max}$ for all $k\in[0,N]$.
For the reference steering angle, we choose $\delta_\mathrm{ref}(k)=0$ for all $k\in[0,N]$. %
Thus, we arrive at the reference trajectory $(x_\mathrm{ref}(k))_{k\in[0,N]}$ with $x_\mathrm{ref}(k) = [p_{\mathrm{ref},x}(k),p_{\mathrm{ref},y}(k),0.9 v_\mathrm{max},\cos(\psi_\mathrm{ref}(k)),\sin(\psi_\mathrm{ref}(k)),0]^\top$.

Before obtaining a new solution to the MPC problem at time $t$, we update the reference trajectory by projecting the estimated e-scooter position to the path and repeating the procedure described above for the updated portion of the path.
To indicate this update of the reference trajectory, we utilize the notation $x_\mathrm{ref}(\cdot \vert t)$ throughout the remainder of this work, referring to the reference trajectory at time $t$.

Moreover, for all $t\geq0$, we define an input reference trajectory $u_\mathrm{ref}(\cdot \vert t)=0$.
We only use this reference to penalize high inputs (with a low weight), as we will see in more detail in the next section.
The state reference tracking performance can in principle be improved by taking a model-based approach for identifying a more suitable $u_\mathrm{ref}(\cdot \vert t)$.
We plan to explore this potential in future research and rely on this simpler approach for this paper, which naturally comes with reduced computational efforts.

\subsection{MPC problem formulation}

We employ an MPC scheme in order to solve the reference trajectory tracking problem resulting from the previous step.
We can formulate the problem with reference $r(t) = [x_\mathrm{ref}(\cdot \vert t), u_\mathrm{ref}(\cdot \vert t)]$ as
\begingroup
\allowdisplaybreaks
\begin{subequations}\label{eq:MPC_for_tracking}
    \begin{equation}\label{eq:MPC_for_tracking_cost}
        W(x(t), r(t)) = \min_{u(\cdot \vert t)} V(x(\cdot \vert t), u(\cdot \vert t), r(t))
    \end{equation}
    subject to
    \begin{alignat}{2}
        x(0 \vert t)                         & = x(t), \label{eq:MPC_IC}
        \\
        x(k+1 \vert t)                       & = f(x(k \vert t), u(k \vert t)),                       & \; k & \in \{0,\dots,N\},
        \label{eq:MPC_for_tracking_dynamics}
        \\
        0                                    & \leq v(k \vert t)            \leq v_\mathrm{max},      & \; k & \in \{0,\dots,N\}, \label{eq:MPC_state_constr1}      \\
        -\delta_\mathrm{max}                                & \leq \delta(k \vert t)       \leq \delta_\mathrm{max},                & \; k & \in \{0,\dots,N\},                                   \\
        -0.4                                 & \leq \dot{\delta}(k \vert t) \leq 0.4,                 & \; k & \in \{0,\dots,N\},                                   \\
        -1.0                                 & \leq a(k \vert t)            \leq 0.7,                 & \; k & \in \{0,\dots,N\}, \label{eq:MPC_input_constr2}      \\
        |\dot{\phi}_\mathrm{cmd}(k \vert t)| & \leq 0.0175,                                           & \; k & \in \{0,\dots,N\}, \label{eq:MPC_balancing_constr1}  \\
        |v(k \vert t)|                       & \leq \frac{v_\mathrm{max}}{1+\mu |\delta(k \vert t)|}, & \; k & \in \{0,\dots,N\}, \label{eq:MPC_balancing_constr2}  \\
        0                                    & \leq \mathrm{sdf}_{\mathrm{f}}(x(k \vert t)),          & \; k & \in \{0,\dots,N\}, \label{eq:MPC_path_width_constr1} \\
        0                                    & \leq \mathrm{sdf}_{\mathrm{r}}(x(k \vert t)),          & \; k & \in \{0,\dots,N\}, \label{eq:MPC_path_width_constr2}
    \end{alignat}
\end{subequations}
\endgroup
with $v_\mathrm{max}=\SI{0.7}{\meter\per\second}$, $\delta_\mathrm{max}=\SI{0.65}{\radian}$, and
\begin{multline*}
    V(x(\cdot \vert t), u(\cdot \vert t), r(t)) = \|x(N \vert t) - x_\mathrm{ref}(N \vert t)\|_P^2 + \\
    \sum_{k=0}^{N-1} \left(\|x(k \vert t)-x_\mathrm{ref}(k \vert t)\|_Q^2 + \|u(k \vert t)-u_\mathrm{ref}(k \vert t)\|_R^2\right).
\end{multline*}
Here, the control inputs $u(k \vert t)$ are the acceleration of the e-scooter at the rear axle $a(k \vert t)$ and the steering angle velocity $\dot{\delta}(k \vert t)$.
Moreover, we use the notation $\|x\|^2_A=x^\top A x$ for any positive definite matrix $A$, and set
\begin{align*}
    Q & = \mathrm{diag}(0.1,0.1,0.04,0.15,0.15,0.0025), \\
    R & = \mathrm{diag}(0.01,0.001),                    \\
    P & = Q.
\end{align*}

In \eqref{eq:MPC_IC}, the state $x(0 \vert t)$ is initialized with the state estimate $x(t)$ obtained from the EKF-based localization at time $t$.

The discrete-time dynamics \eqref{eq:MPC_for_tracking_dynamics} are obtained algorithmically via implicit Runge-Kutta integration of the continuous-time dynamics
\begin{equation*}
    \dot{x} =
    \begin{bmatrix}
        v\cos(\psi) - L\sin(\psi)\dot{\psi}  \\
        v\sin(\psi) + L \cos(\psi)\dot{\psi} \\
        a                                    \\
        -\sin(\psi)\dot{\psi}                \\
        \cos(\psi)\dot{\psi}                 \\
        \dot{\delta}
    \end{bmatrix}
\end{equation*}
with $\dot{\psi} = v \tan(\delta) / L$ as in the kinematic single track model \eqref{eq:single_track_model}.
We utilize \texttt{acados}~\cite{Verschueren2021} to implement the MPC problem.

We deploy box constraints on the e-scooter states and inputs \eqref{eq:MPC_state_constr1}-\eqref{eq:MPC_input_constr2}.
Furthermore, to constrain the inputs of the MPC scheme to be consistent with the low-level balancing controller which filters the commanded velocity and steering angle, we employ nonlinear constraints \eqref{eq:MPC_balancing_constr1}, \eqref{eq:MPC_balancing_constr2}.
This ensures that the low-level balancing controller can maintain the balance of the e-scooter at all times and the MPC prediction does not deviate systematically from the applied low-level reference signals.
In \eqref{eq:MPC_balancing_constr1}, we bound the change rate of roll angle set point $\dot{\phi}_\mathrm{cmd}$ such that the required roll angle motion for balancing is not arbitrarily fast.
As the steady state of the roll dynamics from \cite{Defoort2009}, we obtain (with some simplifications)
\begin{equation*}
    \phi_\mathrm{cmd}(k \vert t) = \arctan\left(\frac{v(k \vert t)^2 \tan \delta(k \vert t)}{Lg}\right),
\end{equation*}
where $g$ is the gravitational acceleration.
This yields
\begin{equation*}
    \dot{\phi}_\mathrm{cmd}(k \vert t) = \frac{Lg(2v\tan(\delta)a + v^2 / \cos^2(\delta) \dot{\delta})}{L^2g^2 + v^4 \tan^2(\delta)},
\end{equation*}
where we omitted the argument $(k \vert t)$ for brevity.
Constraint \eqref{eq:MPC_balancing_constr1} effectively limits the steering angle change rate $\dot{\delta}(k \vert t)$ and the acceleration at the rear axle $a(k \vert t)$, to be compatible with the low-level balancing controller.
The chosen bound $\SI{0.0175}{\radian\per\second}$ is equivalent to $\SI{1}{\degree\per\second}$.
Moreover, constraint \eqref{eq:MPC_balancing_constr2} further reduces the velocity limit in curves in order to facilitate also abrupt stopping maneuvers while balancing, e.g., when an obstacle suddenly crosses the e-scooter's path.
We choose $\mu=(v_\mathrm{max}-v_\mathrm{curve})/(v_\mathrm{curve}\delta_\mathrm{max})$ with $v_\mathrm{curve}=\SI{0.4}{\meter\per\second}$ as the reduced velocity limit in curves with the minimal possible curve radius.

We also have constraints to stay within the path and its width for all time instants \eqref{eq:MPC_path_width_constr1}, \eqref{eq:MPC_path_width_constr2}.
These are also nonlinear constraints in our MPC scheme.
To this end, we define a signed distance function that combines all the line segments in the considered path into one function which is negative if a point on the e-scooter $p=[p_{x}, p_{y}]^\top$ is outside the constraint and positive if $p$ is inside the constraint.
Examples of such points are the front axle position $p_\mathrm{f}=[p_{\mathrm{f},x}, p_{\mathrm{f},y}]^\top$ and the rear axle position $p_\mathrm{r}=[p_{\mathrm{r},x}, p_{\mathrm{r},y}]^\top$, introduced in \eqref{eq:pos_f_trafo}, \eqref{eq:pos_r_trafo}.
In particular, for any $S$ line segments with width $2w_i>0$ for $i=1,\ldots,S$, we define the signed distance function as follows.
Consider a point on the e-scooter $p$ and the $i$-th line segment with start point $a_i$, end point $b_i$, and width $2w_i$.
Then, we define the projection of $p$ onto (the closest point) on each line segment
\begin{equation}\label{eq:proj_sdf}
    h_i(p) = \max\left\{\min\left\{ \frac{(p-a_i)^\top(b_i - a_i)}{\|b_i - a_i\|^2}, 1 \right\}, 0\right\}
\end{equation}
and the signed distance function for each segment
\begin{align*}
    \mathrm{sdf}_{i}(p) & = \frac{w_i^2 - \|p - (1-h_i(p))a_i - h_i(p)b_i\|^2}{w_i^2}.
\end{align*}
The minimization and the maximization in the projection \eqref{eq:proj_sdf} ensure that we only project to points in the line segment.
Consequently, the path width constraints in the MPC scheme are formulated with the signed distance functions $\mathrm{sdf}_\mathrm{f}(x) = \max_{i\in[1,S]}\{\mathrm{sdf}_i(p_{\mathrm{f}})\}$ and $\mathrm{sdf}_\mathrm{r}(x) = \max_{i\in[1,S]}\{\mathrm{sdf}_i(p_{\mathrm{r}})\}$.

Note that we implicitly assume that the reference tracking problem is posed such that a feasible solution exists.
In future work, we plan to explore reference trajectory tracking with an intermediate artificial reference as in~\cite{Limon2008} which can allow to circumvent this assumption on the problem formulation.

As a final step, we transform the inputs of the MPC scheme $u(k \vert t) = [a(k \vert t),\dot{\delta}(k \vert t)]^\top$ into the required reference signals for the low-level controllers shown in Fig.~\ref{fig:cl_arch}.
Given the most recent MPC input trajectory $u(\cdot \vert t)$, we obtain reference signals $[v_\mathrm{cmd}(\tau), \delta_\mathrm{cmd}(\tau)]^\top$ for the low-level controllers, parametrized in time $\tau\geq t$, by integration.

\section{Experiment}
\label{sec:exp}

\begin{figure}
    \centering
    \input{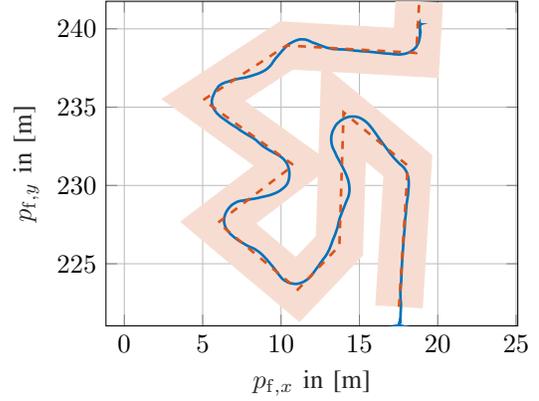}
    \vspace*{-0.5\baselineskip}
    \caption{E-Scooter position (\ref{fig:e-scooter_pos}) and path to follow (\ref{fig:path}) with path width constraints shown as red shaded area.}
    \label{fig:exp_pos}
    \vspace*{-\baselineskip}
\end{figure}

\begin{figure}
    \centering
    \input{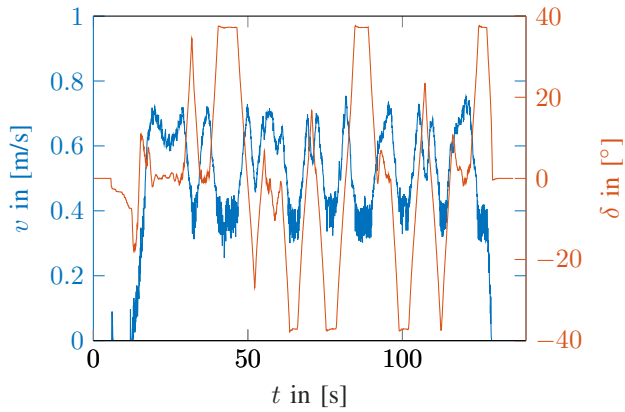}
    \vspace*{-0.5\baselineskip}
    \caption{E-Scooter velocity and steering angle.}
    \label{fig:exp_vel}
    \vspace*{-0.5\baselineskip}
\end{figure}

In this section, we conduct a real-world experiment on our prototype.$^\text{\ref{foot:video}}$
The path to follow for the experiment is shown in Fig.~\ref{fig:exp_pos}, transformed from GNSS coordinates to an East-North-Up (ENU) reference frame for improved interpretability.
The estimated e-scooter position is also depicted.
We observe a satisfying path following performance with respect to the deviation from the path and the progress along the path.
In particular, the e-scooter velocity and steering angle are presented in Fig.~\ref{fig:exp_vel}.
The velocity is reduced with increased absolute steering angle, as prescribed by MPC constraint \eqref{eq:MPC_balancing_constr2}.


\begin{figure}
    \centering
    \input{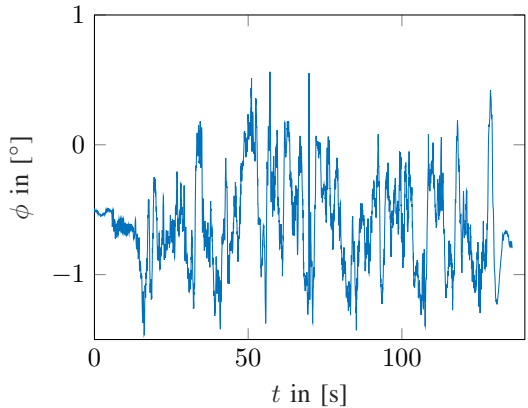}
    \hspace*{0.15\columnwidth}
    \vspace*{-0.5\baselineskip}
    \caption{E-Scooter roll angle.}
    \label{fig:exp_roll}
    \vspace*{-\baselineskip}
\end{figure}

Lastly, Fig.~\ref{fig:exp_roll} displays the estimated roll angle of the e-scooter.
We observe that the e-scooter maintains its balance throughout the experiment.
We also see that our balancing controller does not stabilize roll angle $\phi=0$ and will identify the equilibrium online \cite{Wenzelburger2020}.
Variations in the roll angle are in parts caused by disturbances during operation, e.g., from the ground, or set point changes as the e-scooter is required to lean slightly into the curve.
Note that the roll angle variations however remain in a small regime which demonstrates the efficacy of our closed-loop architecture.

\section{Conclusions}
\label{sec:concl}

In this work, we have proposed a closed-loop architecture to solve the path following problem for an autonomous e-scooter that maintains its balance with a reaction wheel.
We have designed a localization algorithm based on an EKF and an MPC scheme to obtain a solution to the path following problem.
In particular, our approach turns a local portion of the path to follow into a reference trajectory, facilitating a reference tracking MPC scheme to determine feasible reference signals for the low-level controllers.
Furthermore, we have demonstrated the effectiveness of our proposed architecture in a real-world experiment on our prototype.

Our proposed solution combines newly designed high-level planning layers with low-level control for driving, steering, and balancing.
In addition, it is immediately compatible with the collision avoidance safety filter proposed in~\cite{strasser:seidel:braendle:meister:soloperto:hambach-ferrer:allgoewer:2024}.
This is an important step towards fully autonomous e-scooters and other micromobility solutions yet to come.
We plan to explore reference tracking with an artificial reference and direct path following approaches to improve the theoretical properties of our solution.
Moreover, considering the balancing dynamics explicitly in the MPC scheme can further improve the performance and properties of the path following algorithm, for example allowing for higher agility in curves and stability guarantees.
The consideration of detected static and dynamic obstacles in the online planning algorithm is another extension which can be readily integrated in our MPC scheme.
Moreover, we develop a solution to automatically move e-scooters lying on the ground to the balancing state.

\bibliographystyle{IEEEtran}
\bibliography{e-scooter}

\end{document}